\documentclass[aps,prl,twocolumn,groupedaddress,notitlepage,
floatfix,superscriptaddress]{revtex4-2}

\pdfoutput=1 
\usepackage{graphicx,graphics,epsfig,subfigure,times,bm,bbm,amssymb,amsmath,amsthm,mathrsfs,MnSymbol} \usepackage{gensymb} \usepackage{amsfonts} \usepackage{float} \usepackage[matrix,frame,arrow]{xypic} \usepackage[pdfstartview=FitH]{hyperref} 
\usepackage{times}
\usepackage{float}
\usepackage{graphics}
\usepackage[T1]{fontenc}

\usepackage{braket}  
\usepackage{enumerate} \usepackage[normalem]{ulem}
\usepackage[usenames,dvipsnames]{xcolor} \usepackage{multirow} \usepackage{mathtools}
\usepackage{bbm} \usepackage{titletoc} 

\definecolor{orange}{rgb}{1,0.5,0} 
\hypersetup{ colorlinks=true,       
	linkcolor=red,          
	citecolor=blue,        
	filecolor=magenta,      
	urlcolor=blue,           
	runcolor=cyan }

\begin{document}
	\title{Nonmesonic Quantum Many-Body Scars in a 1D Lattice Gauge Theory}

	\author{Zi-Yong~Ge}
	\email{ziyong.ge@riken.jp}
	\affiliation{Theoretical Quantum Physics Laboratory, Cluster for Pioneering Research, RIKEN, Wako-shi, Saitama 351-0198, Japan}

   \author{Yu-Ran~Zhang}
   \email{yuranzhang@scut.edu.cn}
   \affiliation{School of Physics and Optoelectronics, South China University of Technology, Guangzhou 510640, China}

	\author{Franco Nori}
	\email{fnori@riken.jp}
	\affiliation{Theoretical Quantum Physics Laboratory, Cluster for Pioneering Research, RIKEN, Wako-shi, Saitama 351-0198, Japan}
	\affiliation{Quantum Information Physics Theory Research Team, Center for Quantum Computing, RIKEN, Wako-shi, Saitama 351-0198, Japan}
	\affiliation{Department of Physics, University of Michigan, Ann Arbor, Michigan 48109-1040, USA}

	\begin{abstract}
	We investigate the meson excitations (particle-antiparticle bound states) in quantum many-body scars 
	of a 1D $\mathbb{Z}_2$ lattice gauge theory coupled to a dynamical spin-$\frac{1}{2}$ chain as a matter field.
	By introducing a string representation of the physical Hilbert space,
	we express a scar state $\ket {\Psi_{n,l}}$ as a superposition of all string bases with an identical string number $n$ and a total length $l$.
	For the small-$l$ scar state $\ket {\Psi_{n,l}}$, the  gauge-invariant spin exchange correlation function of  the matter field
	hosts an exponential decay as the distance increases,
	indicating the existence of stable mesons.
	However, for large $l$, the correlation function exhibits a power-law decay, signaling the emergence of  nonmesonic excitations.
	Furthermore, we show that this mesonic-nonmesonic crossover can be detected by the quench dynamics, starting from two low-entangled initial states, respectively,
	which are experimentally feasible in quantum simulators.
	Our results expand the physics of quantum many-body scars in lattice gauge theories
	and reveal that the nonmesonic state can also manifest ergodicity breaking.
\end{abstract}

\maketitle

\textit{Introduction.}---%
Because of the development of quantum simulations~\cite{doi:10.1080/00018730701223200,RevModPhys.80.885,Buluta108,Buluta_2011,doi:10.1080/00107514.2016.1151199, RevModPhys.86.153,Gross995,s41567-019-0733-z},  out-of-equilibrium quantum many-body physics has been attracting growing interests~\cite{RevModPhys.83.863}.
The Eigenstate Thermalization Hypothesis (ETH) postulates that generic isolated nonintegrable quantum many-body systems exhibit ergodicity~\cite{PhysRevA.43.2046,PhysRevE.50.888,Srednicki_1999,Rigol2008Nature,DAlessio2016},
and thus, the unitary quantum evolution of the systems can result in an equilibrium state described by  statistical mechanics.
Though ETH was thought to be  general, there are several counterexamples,
e.g., quantum integrable systems~\cite{kinoshita2006quantum,PhysRevLett.98.050405} and many-body localizations~\cite{Basko2006,PhysRevB.82.174411,PhysRevLett.113.107204,PhysRevB.90.174202,PhysRevB.93.041424,Rahul2015,RevModPhys.91.021001}.
These two examples are called strong ergodicity breaking, since most of the eigenstates violate the ETH.
Recent experimental and theoretical works demonstrate that there exists a new type of ETH-violating eigenstates in some specific nonintegrable quantum many-body systems,  dubbed quantum many-body scar (QMBS) states~\cite{bernien2017probing,turner2018weak,PhysRevB.98.155134,PhysRevLett.122.173401,PhysRevB.98.235155,PhysRevB.98.235156,PhysRevLett.122.220603,PhysRevLett.122.040603,Shiraishi_2019,PhysRevLett.123.147201,PhysRevResearch.2.033044,PhysRevB.102.041117,PhysRevB.101.195131,PhysRevB.102.075132,PhysRevB.102.085140,PhysRevResearch.2.043305,PhysRevLett.124.180604,PhysRevLett.125.230602,PhysRevLett.126.120604,Moudgalya_2022,PhysRevResearch.5.023010}.
Generally, the number of QMBS states is  exponentially smaller than the Hilbert space dimension,
so they can be considered as  weak ergodicity breaking.
One typical class of QMBS states is constructed by spectrum-generated algebras~\cite{buvca2019non,PhysRevB.102.085140,PhysRevB.102.075132,Moudgalya_2022},
whose eigenenergies  are equally spaced, dubbed towers of QMBSs~\cite{Moudgalya_2022}.
Thus, if the initial state is a superposition of these scar states, there will exist
a perfect revival dynamics indicating the ETH violation.

Empirically, kinetically constrained systems are thought more likely to host QMBSs.
Thus, as a typical instance, lattice gauge theories (LGTs)~\cite{PhysRevD.10.2445,RevModPhys.51.659,PhysRevB.65.024504,PhysRevD.17.2637,PhysRevD.19.3682,Fradkin2013,PhysRevA.73.022328,
	PhysRevLett.109.175302,di2019resolution,Magnifico2020realtimedynamics,banuls2020simulating,
	PhysRevResearch.3.023079,PhysRevA.103.053703,PRXQuantum.3.010324,Halimeh2023robustquantummany,PhysRevB.107.L201105,PhysRevB.107.205112}
have attracted considerable interest to study QMBSs~\cite{PhysRevLett.122.130603, PhysRevB.99.195108, PhysRevX.10.021041,
	PhysRevB.101.024306,PhysRevLett.124.207602,PhysRevLett.126.220601,PhysRevB.106.L041101,PhysRevResearch.4.L032037}.
Meanwhile, quasiparticles are significant for understanding the towers of QMBSs~\cite{doi:10.1146/annurev-conmatphys-031620-101617}.
In LGTs, particles can be pairwise confined into mesons, which are one of the most important quasiparticles,
closely related to the towers of QMBS~\cite{PhysRevLett.122.130603, PhysRevB.99.195108, PhysRevX.10.021041,
	PhysRevB.101.024306, ISI:000395814000014,birnkammer2022prethermalization}.
For instance, the scar dynamics in the \textit{PXP} model~\cite{PhysRevX.10.021041} can be understood
as the string inversion of a $U(1)$ LGT,
where a particle and an antiparticle form a bound state, a stable meson.
Therefore, one natural question is whether stable mesons  are necessary conditions for the towers of QMBSs in LGTs.
A recent work~\cite{PhysRevB.101.195131} reports a special type of QMBSs in a spin chain with conserved  domain wall (equivalent to a $\mathbb{Z}_2$ LGT),
which is generated by nonlocal operators.
Based on these QMBSs, we address the above questions, and further reveal more
nontrivial physics when investigating QMBSs in LGTs.

In this Letter, we study this type of QMBSs in a $\mathbb{Z}_2$ LGT
and demonstrate that it can manifest both \textit{mesonic} and \textit{nonmesonic} features.
First, we introduce a string representation to describe the physical Hilbert space in a specific gauge sector.
In this representation, the exact wave function of the QMBS state $\ket {\Psi_{n,l}}$ is written as an equal superposition of
all string bases with an identical string number $n$ and a total string length $l$.
We identify the meson properties in $\ket {\Psi_{n,l}}$ by calculating the  gauge-invariant spin exchange correlation function of the matter field.
Our results show that there exist stable mesons for small $l$, while the quasiparticles are  nonmesonic for large $l$.
Moreover, we propose a feasible approach to demonstrate the mesonic-nonmesonic crossover with quantum simulators,
by observing the quench dynamics of the system initially at two different low-entangled states, respectively.

\textit{Model.}---%
Here, we consider a 1D $\mathbb{Z}_2$ LGT minimally coupled to  a dynamical spin-$\frac{1}{2}$ chain as a matter field.
The  Hamiltonian has a form $\hat H = \hat H_K + H_E + \hat H_\mu$, with 
\begin{align} \label{H} \nonumber
	&\hat  H_K =-J\sum_{j=1}^L\big(\hat \sigma^+_{j}\hat \tau^z_{j+\frac{1}{2}}\hat \sigma^-_{j+1,} + \text{H.c.}\big),  \\
	&\hat H_E = -h\sum_{j=1}^L \hat \tau_{j+\frac{1}{2}}^x,\ \ \ \  \hat H_\mu = \mu\sum_{j=1}^L \hat \sigma^+_{j}\hat \sigma^-_{j},
\end{align}
where $\hat \sigma^\alpha_{j}$ and  $\hat \tau_{j+\frac{1}{2}}^\alpha$ are both Pauli matrices describing the matter and gauge fields, respectively,
and $L$ is the system size. 
The matter field lives on the site, while the gauge field is on the link.
The kinetic term $\hat H_K$ describes the minimal  gauge-matter coupling with strength $J$,
the second term $\hat H_E$ describes an electric field with strength $h$, and
the last term $\hat H_\mu$ denotes the potential of the matter field.
We consider periodic boundary conditions, i.e., $\hat \sigma^\alpha_{1}=\hat \sigma^\alpha_{L+1}$ and $\hat \tau_{1+\frac{1}{2}}^\alpha=\hat \tau_{L+1+\frac{1}{2}}^\alpha$.
The Hamiltonian $\hat H$ is $\mathbb{Z}_2$ gauge invariant with a generator
$\hat G_j = \hat \tau_{j-\frac{1}{2}}^x \hat\sigma^z_j\hat\tau_{j+\frac{1}{2}}^x$.
In addition to the gauge structure, there also exists the spin $U(1)$ symmetry in $\hat H$, where the total spin charge $\sum_{j=1}^L \hat \sigma^+_{j}\hat \sigma^-_{j}$ is conserved.
Note that this $\mathbb{Z}_2$ LGT can be experimentally addressed in various quantum simulators~\cite{ Barbieroeaav7444,ISI:000494944200024,ISI:000494944200023,
	Ge_2021,PhysRevResearch.4.L022060,arXiv:2203.08905,arXiv:2206.08909,PhysRevB.107.125141}.

Without loss of generality, we fix the system to the gauge sector $\hat G_j = 1$,
with an even number of total spin charges for each physical basis.
With a dual transformation,
the original Hamiltonian $\hat H$ can be mapped to a local spin chain~\cite{PhysRevLett.127.167203}
\begin{align} \label{Hdual}\nonumber
	&\hat H_K = -\frac{J}{2}\sum_{j}(\hat Z_{j+\frac{1}{2}}-\hat X_{j-\frac{1}{2}}\hat Z_{j+\frac{1}{2}}\hat X_{j+\frac{3}{2}}),\\
	&\hat H_E = -h\sum_{j}\hat X_{j+\frac{1}{2}}, \ \ \ \ \hat H_\mu =\frac{\mu}{2}\sum_{j}(1-\hat X_{j-\frac{1}{2}}\hat X_{j+\frac{1}{2}}),
\end{align}
where $\hat X_{j+\frac{1}{2}}=\hat \tau^x_{j+\frac{1}{2}}$ and $\hat Z_{j+\frac{1}{2}}
=\hat\sigma^x_j \hat \tau^z_{j+\frac{1}{2}}\hat\sigma^x_{j+1}$
are also Pauli matrices.

\begin{figure}[t] \includegraphics[width=0.47\textwidth]{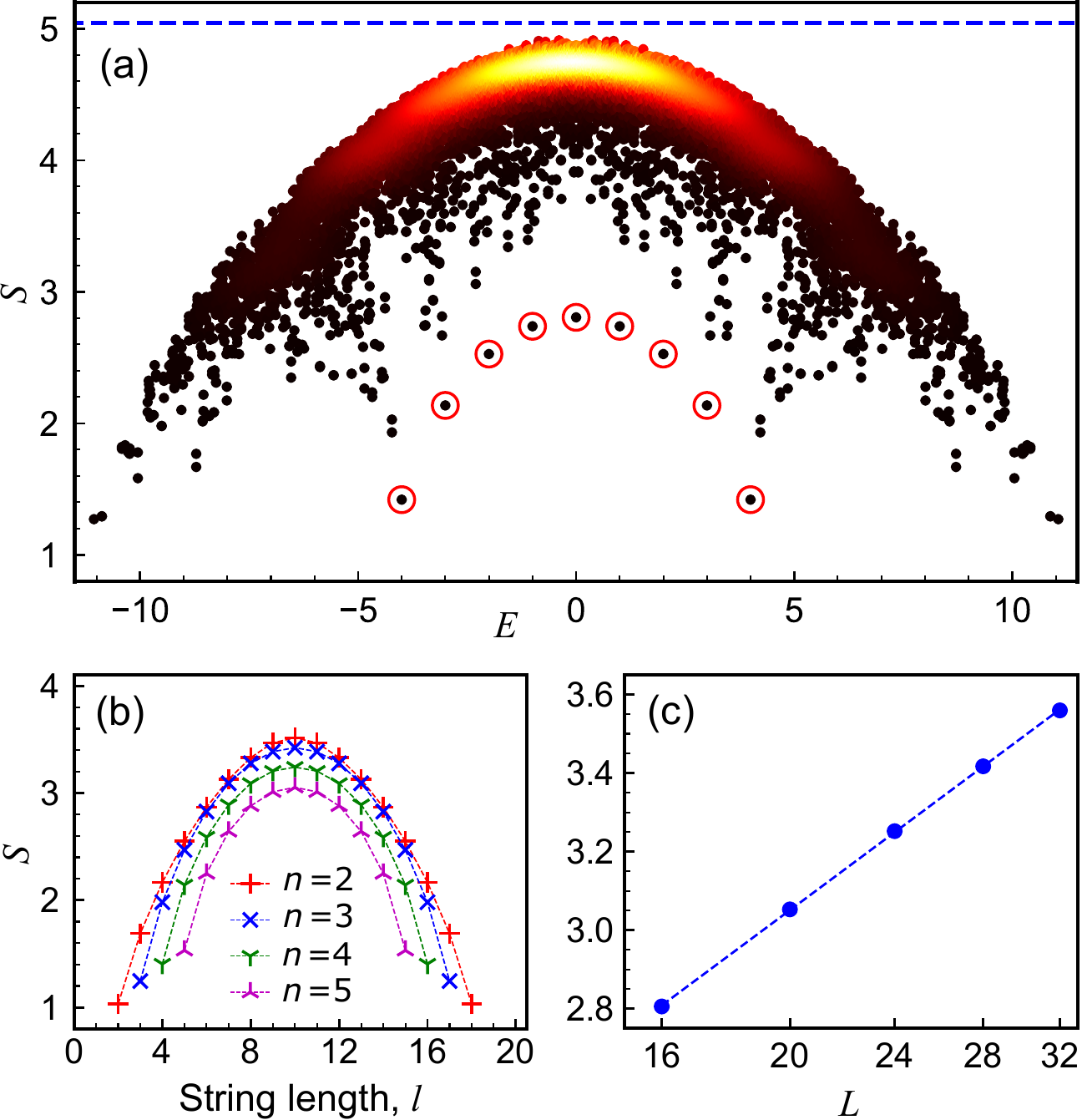}
	\caption{Half-chain von Neumann  entropy $S$.
		(a) The distribution of $S$ for all eigenstates of Hamiltonian (\ref{H}) in the half-filling sector (i.e., string number $n=L/4$) with $L=16$, $J=1$, $h=0.5$.
		The color codes the density of states (warmer colors imply higher density).
		The blue dashed line represents the entanglement entropy of the random state $S^{\text{ran}}=(L\ln 2-1)/2\approx5.05$.
		The data points in red circles correspond to the eigenstates in Eq.~(\ref{psi_nl}).
		(b) Pyramidlike structure of $\ket {\Psi_{n,l}}$ for $L=20$.
		Different colors represent different numbers of strings $n$.
		(c) The size scaling of the entanglement entropy for the QMBS state $\ket{\Psi_{L/4,L/2}}$. The dashed line shows linear fitting: $S\sim \ln L$.}
	\label{fig_1}
\end{figure}

Figure~\ref{fig_1}(a) plots the half-chain von Neumann entropies, 
$S=\text{Tr}\hat\rho_{L/2}\ln \hat\rho_{L/2}$, of the whole eigenstates in the half-filling case ($\sum_{j=1}^L \hat \sigma^+_j\hat \sigma^-_j=L/2$),	
where $\hat\rho_{L/2}$ is the half-chain density matrix.
The entanglement entropies of states near the middle of the spectrum
approach the value for a random	state $S^{\text{ran}}=(L\ln 2-1)/2$~\cite{PhysRevLett.71.1291},
which demonstrates that most of the eigenstates obey ETH.

As shown in  Ref.~\cite{PhysRevB.101.195131}, a pyramid-like structure of scar states exists for the dual Hamiltonian (\ref{Hdual}), see Figs.~\ref{fig_1}(a,b).
These scar states cannot be generated by local operators and are very distinct from the conventional towers of QMBSs.
However, they have not been fully investigated. Specifically, 
it is still unclear whether stable meson excitations dominate the scar dynamics, like most of conventional QMBSs in LGTs.
In addition, an experimental proposal to detect these QMBSs from the nontrivial quench dynamics
in quantum simulators is also a relevant issue.
Hereafter, we investigate the QMBSs of $\hat H$ from the viewpoint of LGTs,
and uncover whether these can be described by mesonic physics.
Note that while analytical discussions are based on the original Hamiltonian Eq.~(\ref{H}),
the numerical results are obtained from the Hamiltonian (\ref{Hdual}) by exact diagonalization.

\textit{String representation.}---%
We introduce string bases in the $\hat G_j = 1$ sector,
which are convenient for discussing meson excitations.
Because of gauge invariance, the physical Hilbert space in the fixed gauge sector can be represented by open strings.
The vacuum state of string excitations can be defined as $\ket{\Omega} :=\ket{\downarrow\downarrow...\downarrow\downarrow}\otimes\ket{++...++}$,
where $\ket{\downarrow\downarrow...\downarrow\downarrow}$ is a ferromagnetic state of matter fields,
and $\ket{\tau}=\ket{++...++}$ is the state of gauge fields with all links being polarized at $\hat \tau^x=1$.
A state with one string excitation can be written as
\begin{align} \label{Skl}
	\ket{\mathcal{S}_{k,\ell}}  :=\hat {\mathcal{S}}_{k,\ell}^\dagger\ket{\Omega}=\hat \sigma_k^+ \big(\prod_{k\leq j<k+\ell} \hat \tau^{\text{M}}_{j+\frac{1}{2}}\big)\hat \sigma_{k+\ell}^+\ket{\Omega},
\end{align}
where $ \hat \tau^{\text{M}}_{j+\frac{1}{2}}=\ket{-}\bra{+}$, and
$k$ denotes the string position, and $\ell$ denotes the string length.
Note that, if the operator $\hat {\mathcal{S}}_{k,\ell}^\dagger$ is local [i.e., $\ell\sim O(1)$], 
the corresponding string excitation can be regarded as a meson.

An arbitrary gauge invariant basis can be written as
\begin{align} \label{vstr}
	\ket{\{\mathcal{S}_{k_j,\ell_j }\}_n^l }  =\hat {\mathcal{S}}_{k_1,\ell_1}^\dagger\hat {\mathcal{S}}_{k_2,\ell_2}^\dagger...
	\hat {\mathcal{S}}_{k_{n-1},\ell_{n-1}}^\dagger\hat {\mathcal{S}}_{k_n,\ell_n}^\dagger\ket{\Omega},
\end{align}
where $k_j > k_i + \ell_i$, for $j>i$, $n$ is the number of strings and equals half of  total spin charges, $l:= \sum_{j=1}^{n} \ell_j$ is the total string length,
and the parity is defined as $P_{\{\mathcal{S}_{k_j,\ell_j }\}_n^l }=\exp{(-i\pi \sum_{j=1}^n k_j)}$.
While $l$ determines the energy of the electric-field term $\hat H_E\ket{\{\mathcal{S}_{k_j,\ell_j }\}_n ^l} = h(2l-L)\ket{\{\mathcal{S}_{k_j,\ell_j }\}_n^l} $,
$n$ determines the energy of the potential term $\hat H_\mu\ket{\{\mathcal{S}_{k_j,\ell_j }\}_n ^l} = 2\mu n\ket{\{\mathcal{S}_{k_j,\ell_j }\}_n^l} $.

\textit{Exact quantum many-body scars.}---%
We introduce the pyramidlike QMBS states in this $\mathbb{Z}_2$ LGT~\cite{PhysRevB.101.195131}, whose wave functions in the string representation are written as
\begin{align} \label{psi_nl}
	\ket {\Psi_{n,l}} = \mathcal{N}_{n,l}\sum_{\{k_j,\ell_j\}} P_{\{\mathcal{S}_{k_j,\ell_j }\}_n^l } \ket{\{\mathcal{S}_{k_j,\ell_j }\}_{n}^{l} },
\end{align}
where $\mathcal{N}_{n,l}$ is a normalization factor.
That is, $\ket {\Psi_{n,l}} $ is an equal superposition of all string bases with both the same string number $n$ and total length $l$,
and the phase is determined by the parity of each basis.
Since $0<\ell_j <L$, the quantum numbers $n$ and $l$ satisfy $n\leq l \leq L-n$.
Thus, there are $(L-2n+1)$ of eigenstates in the sector with $\sum_{j=1}^L\hat\sigma_j^+ \hat\sigma_j^-=2n$ total spin charge.
It can be proved that $\hat H_K \ket {\Psi_{n,l}} = 0$ [see more details in Supplemental Material (SM)~\cite{SM}],
and the eigenenergy of $\ket {\Psi_{n,l}} $ is  ${\varepsilon}_{n,l} = 2hl+2\mu n-hL $,
which can be away from edges of the spectrum corresponding to a high-energy eigenstate, see Fig.~\ref{fig_1}(a).
In addition, the scar states $\ket {\Psi_{n,l}}$ host the sub-volume-law entanglement entropy,
i.e., $S\sim \ln L$, demonstrating the ETH violation, see Fig.~\ref{fig_1}(c).

The scar state  $\ket {\Psi_{n,l}}$ can also be expressed in terms of generating operators.
First, we consider a simple case $\ket {\Psi_{n,n}}$, which only contains $n$ length-$1$ string excitation.
We can construct a ladder operator~\cite{PhysRevB.101.024306}
$ \hat S^\dagger := \sum_{j} P_{\mathcal{S}_{j,1}} \hat{\mathcal{S}}_{j,1}^\dagger$, 
and the eigenstate $\ket {\Psi_{n,n}}$  can be obtained as
\begin{align} \label{psinn}
	\ket {\Psi_{n,n}} = \mathcal{A}_n (\hat S^\dagger) ^n\ket{\Omega},
\end{align}
where $\mathcal{A}_n$ is a normalization factor.
Then, we introduce another operator~\cite{PhysRevB.101.195131} $\hat L^\dagger_m =\sum_{j}  \big(\sum_{k\leq m}\prod_{\ell \leq k}\hat {\mathcal{P}}^-_{j+\frac{1}{2}-\ell} \big)\hat \sigma^-_{j}\hat \tau^{\text{M}}_{j+\frac{1}{2}}\hat \sigma^+_{j+1}$,
where $\hat {\mathcal{P}}^- := \ket{-}\bra{-}$.
The action of $\hat L^\dagger_m$ is to enlarge the total string length by 1 without changing the parity.
Using $\hat L_m^\dag$, we obtain the eigenstate $\ket {\Psi_{n,n+m}}$ as~\cite{PhysRevB.101.195131,SM}
\begin{align} \label{psi2}
	\ket {\Psi_{n,n+m}}  = \mathcal{D}_{n,m}\hat L_m^\dag \ket {\Psi_{n,n+m-1}}
\end{align}
where $\mathcal{D}_{n,m}$ is a normalization factor.
Equations~(\ref{psinn}, \ref{psi2}) indicate that the scar state $\ket {\Psi_{n,n}}$, like the conventional tower of QMBSs, is generated by \textit{local} operators,
while  $\ket {\Psi_{n,n+m}}$ is generated by \textit{nonlocal} operators.
Thus, intuitively, as the total string length $l$ increases, the meson properties for the scar state $\ket {\Psi_{n,l}}$ are expected to be significantly changed.

\begin{figure}[t] \includegraphics[width=0.48\textwidth]{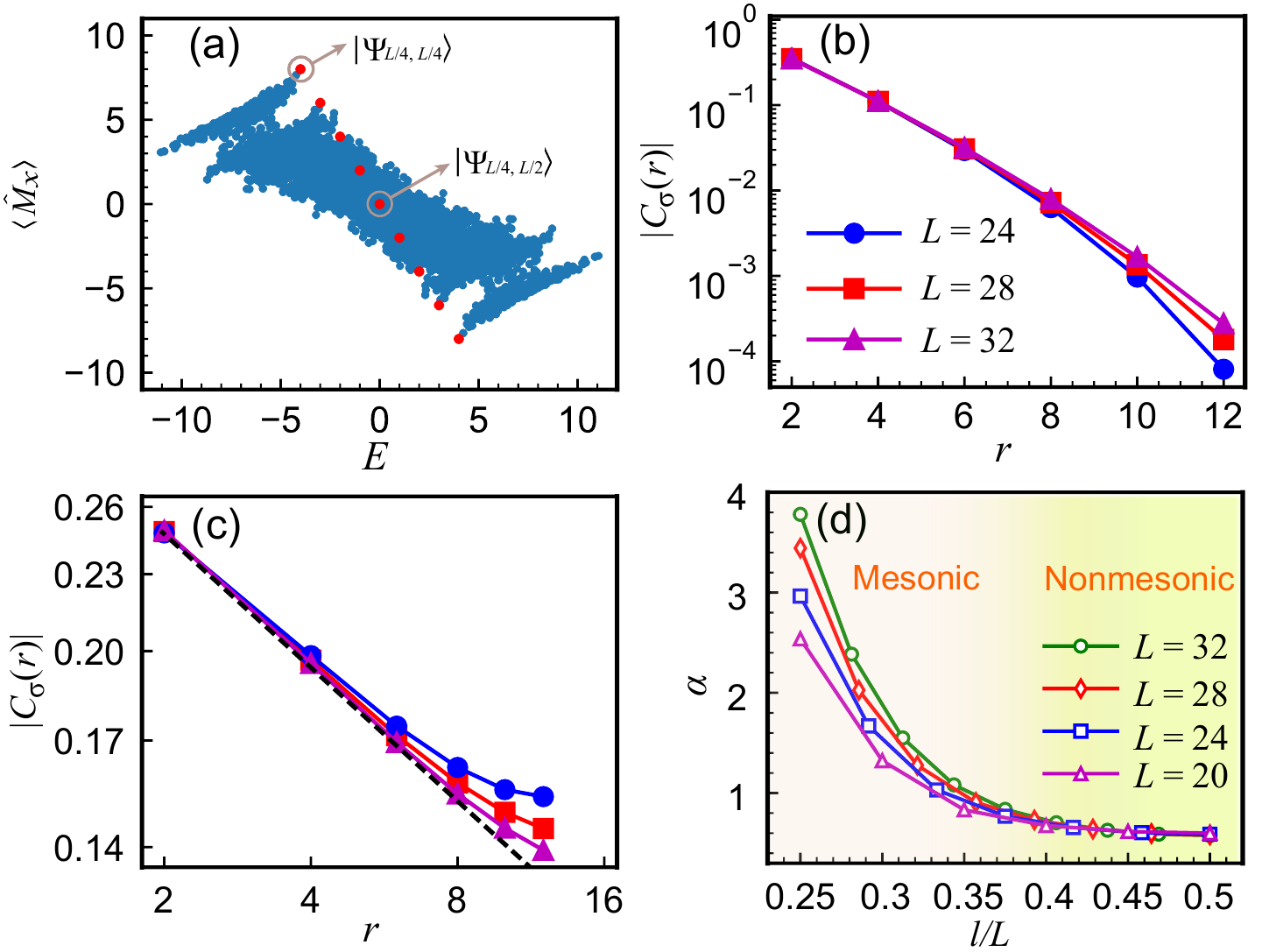}
	\caption{ Properties of mesons. (a) Expectation values of the electric field for all eigenstates at half filling with $L=16$ and $h=0.5$.
		The red dots correspond to scar states. 
		(b,c) Gauge-invariant spin exchange correlation function of the matter field $C_{\sigma}(r)$
		defined in  Eq.~(\ref{Cf}) for $l=L/4$ and $l=L/2$, respectively. The total string number is $n=L/4$.
		The black dashed line is for the fit: $|C_{\sigma}(r)|\sim r^{-\Delta}$, with $\Delta\approx0.35$.
		(d) Ratio $\alpha$ in Eq.~(\ref{alpha}) versus $l/L$ for the scar state $\ket{\Psi_{n,l}}$ with $n=L/4$.}
	\label{fig_2}
\end{figure}

\textit{Mesonic-nonmesonic crossover.}---%
Mesons, as a type of particle-antiparticle bound states, play an important role in the dynamics of LGTs.
If the system is in a confined phase, the low-energy excitation is described by mesons.
In addition, in a high-energy regime, meson dynamics also closely relate to the ETH.
Previous works have shown that almost all of the towers of QMBSs in LGTs originate from meson excitations~\cite{PhysRevLett.122.130603, PhysRevB.99.195108, PhysRevX.10.021041,PhysRevB.101.024306, ISI:000395814000014,birnkammer2022prethermalization}.
For the Hamiltonian (\ref{H}), the $\sigma$-spin is confined in the ground state with an arbitrary finite $h$~\cite{PhysRevLett.127.167203},
where the low-energy excitation is a meson. 
However, it is still unclear whether mesons can describe the high-energy dynamics, especially the scar dynamics.

According to Eq.~(\ref{vstr}), the $\sigma$-spin charges appear pairwise forming string excitations.
For small $l$, e.g., $\ket{\Psi_{n,n}}$, two $\sigma$-spin charges are always bonded together on two nearest-neighbor sites,
i.e., there only exist \textit{local} string excitations (mesons).
This suggests that these scar states should be described by stable mesons.
However, as  $l$ increases, the distance between two $\sigma$ spin charges of a string excitation becomes large, and \textit{nonlocal} string excitations can emerge.
Hence, intuitively, isolated $\sigma$-spin charges are expected to exist in this case, i.e., nonmesonic quasiparticles emerge.
In Fig.~\ref{fig_2}(a), we present the expectation value of the electric field, defined as $\hat M_x:=\sum_{j=1}^L\hat\tau_{j+\frac{1}{2}}^x$.  
It shows that $\braket{\hat M_x}$ of small-$l$ scar states are located at the edge of the spectrum,
which is similar to the confinement-induced non-thermal states in Refs.~\cite{PhysRevLett.122.130603, PhysRevB.99.195108}.
This suggests that these scar states can be described by stable mesons.
However, for large-$l$ scar states, $\braket{\hat M_x}$ is located in the main spectrum,
implying nonmesonic features.
Note that we only need to consider the total string length $l\leq L/2$, and the large total string length means that $l$ is $\approx L/2$~\cite{StringLength}.

To further verify the above picture, 
we perform numerical simulations by calculating the gauge-invariant  spin exchange correlation function versus the distance $r$~\cite{PhysRevLett.127.167203}
\begin{align} \label{Cf}
	C_{\sigma}(r):=\bra{\Psi_{n,l}}\big[ \hat \sigma_j^+(\prod_{j\leq k < j+r} \hat \tau^z_{k+\frac{1}{2}}) \hat \sigma^-_{j+r}+h.c.\big]\ket{\Psi_{n,l}},
\end{align}
which identifies elementary excitations.
Here, if $C_{\sigma}(r)$ exhibits an exponential decay for increasing $r$, 
the isolated $\sigma$-spin charges cannot be detected. Thus, the quasiparticles should be composite particles of matter fields, i.e., mesons~\cite{PhysRevLett.127.167203}.
However, if $C_{\sigma}(r)$ exhibits a power-law decay or converges to a nonzero value with increasing $r$,
an isolated $\sigma$-spin charge can be observed, and the quasiparticles are not mesons.
Figure~\ref{fig_2}(b) shows  that $C_{\sigma}(r)$ approximately exhibits an exponential decay for $\ket{\Psi_{L/4,L/4}}$,
indicating that the small-$l$ scar states are described by stable mesons.
However, $C_{\sigma}(r)$ can exhibit a power-law decay for the state $\ket{\Psi_{L/4,L/2}}$,
showing the existence of nonmesonic excitations for large $l$, see Fig.~\ref{fig_2}(c).
Figure~\ref{fig_2}(d) plots the numerical results of the ratio 
\begin{align} \label{alpha}
	\alpha:=-\ln C_\sigma(L/2)/\ln L.
\end{align}
We find that, as $l/L$ increases, the curves of $\alpha$ for different system sizes gradually collapse to a single curve,
showing that $C_{\sigma}(r)$ gradually exhibits a power-law decay.
Therefore, there exists a mesonic-nonmesonic crossover for these QMBSs.
Figure \ref{fig_2} also reveals that the stable meson excitation is not a necessary condition for the towers of QMBSs in LGTs.
Note that the nonmesonic excitation is a collective effect, which only emerges in many-particle systems, 
i.e., the filling factor $n/L$ is finite.

\begin{figure}[t] \includegraphics[width=0.48\textwidth]{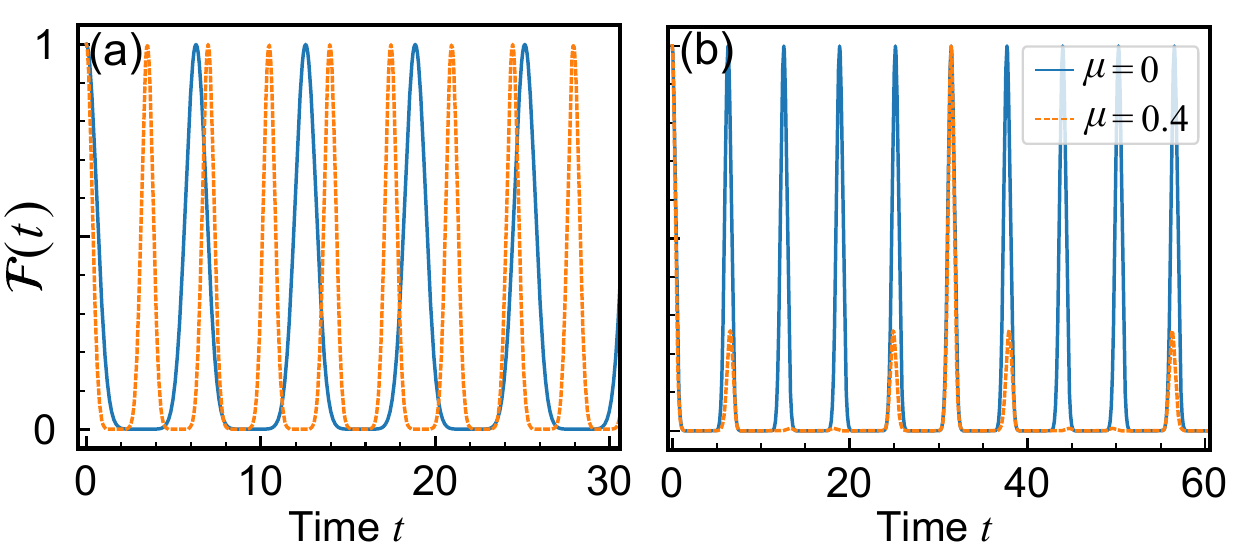}
	\caption{Time evolution of the fidelity  $\mathcal{F}(t)$ after a quantum quench for the initial states (a) $\ket{\psi_1}$ and (b) $\ket{\psi_2}$ in Eq.~(\ref{psi0}).
		Here, we choose $J=1$, $h=0.5$, and $L=16$. }
	\label{fig_3}
\end{figure}

\textit{Quench dynamics.}---%
Another problem is whether the QMBS states $\ket{\Psi_{n,l}}$ can lead to nontrivial quench dynamics,
which can be experimentally observed in quantum simulators.
Here, we introduce two initial states
\begin{subequations}\label{psi0}
	\begin{align} 
		&\ket{\psi_{1} }= \mathcal{B} \prod_{j} \big[1+(-1)^j\hat \sigma^+_{j} \hat \tau^{\text{M}}_{j+\frac{1}{2}}\hat \sigma^+_{j+1}\big]\ket{\Omega},\\
		&\ket{\psi_{2} }= \frac{1}{2^{L/2}} \sum_{n,l}\sum _{\{k_j,\ell_j\}} P_{\{\mathcal{S}_{k_j,\ell_j }\}_n^l } \ket{\{\mathcal{S}_{k_j,\ell_j }\}_{n}^{l} },
	\end{align}
\end{subequations}
where $\mathcal{B}$ is a normalization factor.
Here, $\ket{\psi_{1} }$ is a superposition of scar states $\ket{\Psi_{n,n}}$, i.e., $\ket{\psi_{1} }=\sum_n \alpha_n\ket{\Psi_{n,n}}$,
and $\ket{\psi_{2} }$ is a superposition of all scar states $\ket{\Psi_{n,l}}$, i.e., $\ket{\psi_{2} }=\sum_{n,l} \beta_{n,l}\ket{\Psi_{n,l}}$.
It is obvious that both initial states host low entanglement entropies. Specifically, with the dual transformation in Eq.~(\ref{Hdual}), $\ket{\psi_{1}}$ is related to the ground state of the \textit{PXP} model~\cite{PhysRevB.101.024306}, and $\ket{\psi_{2}} =\bigotimes_j \ket{V_{2j+\frac{1}{2},2j+\frac{3}{2}}}$, with $\ket{V_{2j+\frac{1}{2},2j+\frac{3}{2}}}=(\ket{++}-\ket{+-}+\ket{-+}+\ket{--})/2$.

Figure~\ref{fig_3} presents the fidelity $\mathcal{F}(t):=|\braket{\psi_{1,2}|e^{-i\hat H t}|\psi_{1,2}}|^2$.
It shows that $\mathcal{F}(t)$ exhibits perfect revival dynamics for both initial states in Eq.~(\ref{psi0}).
For the initial state $\ket{\psi_{1} }$, the oscillation period is $T=\pi/(h+\mu)$, see Fig.~\ref{fig_3}(a).
For the initial state $\ket{\psi_{2} }$, if $h/\mu = p/q$, with $p$ and $q$ being relatively prime,
the time for a perfect revival is $T=p\pi/h=q\pi/\mu$, see Fig.~\ref{fig_3}(b).
The oscillation period is consistent with the eigenenergies of $\ket{\Psi_{n,l}}$.
Moreover, the revival dynamics signals the ETH violation for QMBSs  $\ket{\Psi_{n,l}}$.

\begin{figure}[t] \includegraphics[width=0.48\textwidth]{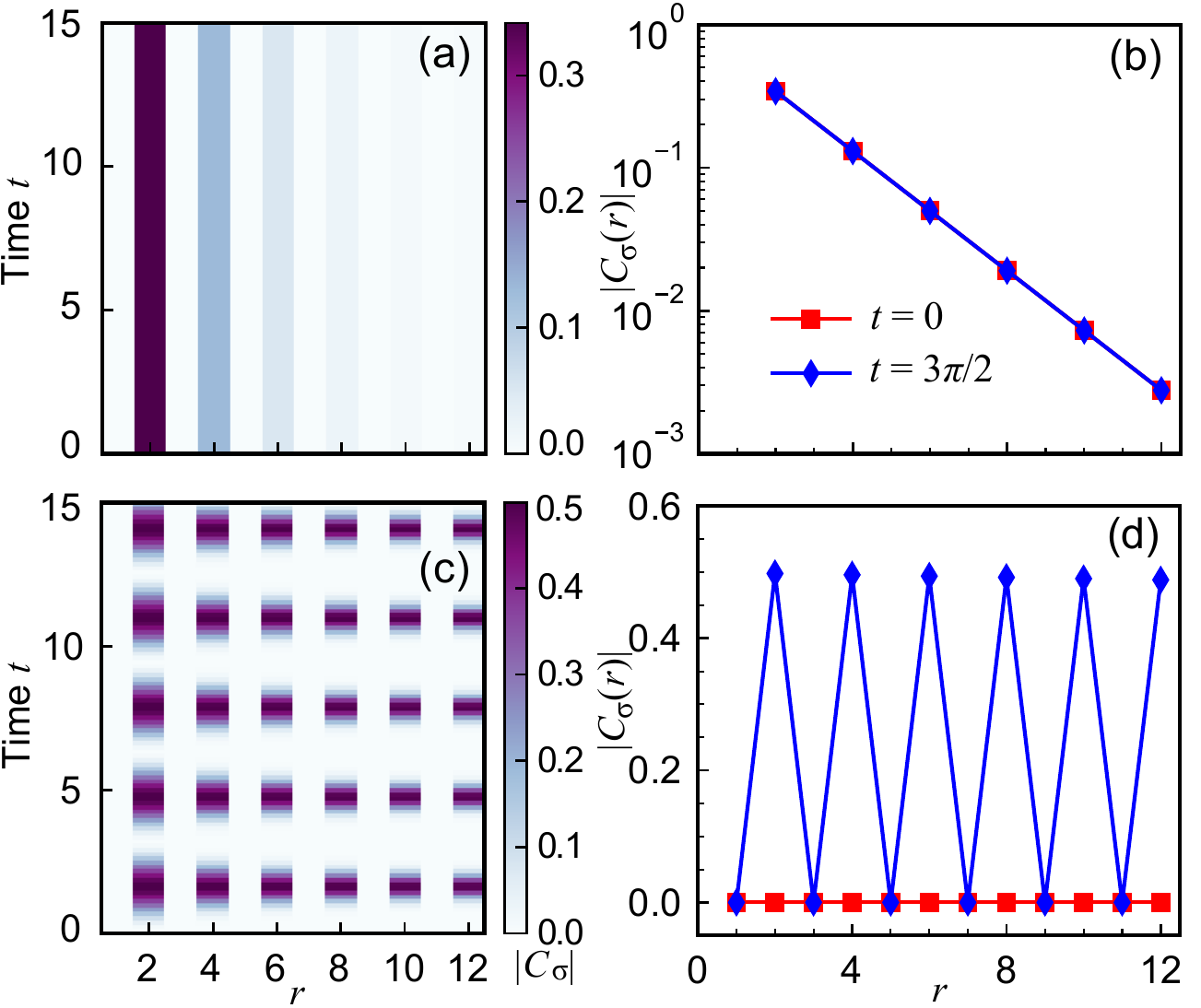}
	\caption{Time evolution of the correlation function $C_{\sigma}(r)$ for two
		different initial states (a,b) $\ket{\psi_1}$ and (c,d) $\ket{\psi_2}$.
		Here, we choose $J=1$, $h=0.5$, $\mu=0.4$, and $L=24$. }
	\label{fig_4}
\end{figure}

We also probe the quasiparticles during the quench dynamics.
Since $\ket{\psi_{1} }$ is a superposition of small-$l$ scar states, we expect a mesonic quench dynamics,
i.e., mesons are always stable during the dynamics, like the scar dynamics in the \textit{PXP} model~\cite{PhysRevX.10.021041}.
For the initial state $\ket{\psi_{2} }$, although it is a superposition of all scar states,
the large-$l$ scar states should be dominant~\cite{SM}, e.g., $|\beta_{L/4,L/2}|\gg |\beta_{L/4,L/4}|$.
Thus, it leads to nonmesonic dynamics.
We calculate the gauge invariant correlation function $C_{\sigma}(r)$ in Eq.~(\ref{Cf})
to identify meson properties during the quench dynamics.
For $\ket{\psi_{1} }$, we find that $C_{\sigma}(r)$ exhibits an exponential decay during the quench dynamics, see Figs.~\ref{fig_4}(a,b).
This indicates that mesons are very stable and cannot be decomposed into isolated spin charges.
However, the situation becomes different for  $\ket{\psi_{2} }$,
where $C_{\sigma}(r)\sim \text{const}$ for $r\rightarrow \infty$ at the specific time, see Figs.~\ref{fig_4}(c,d).
Thus, the isolated spin charges dominate the dynamics, without stable mesons.
Therefore, the initial states in Eq.~(\ref{psi0}) can be used to detect the mesonic-nonmesonic crossover for QMBSs $\ket{\Psi_{n,l}}$
during their quench dynamics.

\textit{Experimental proposal.}---%
The preparation of the initial states in Eq.~(\ref{psi0}) in dual systems  is convenient in quantum simulators.
In addition, the dual Hamiltonian (\ref{Hdual}), with three-body interactions, has also been realized with quantum gates, e.g, in superconducting circuits~\cite{arXiv:2203.08905,zhang2022digital}.
Therefore, the mesonic-nonmesonic crossover for QMBSs in this $\mathbb{Z}_2$ LGT can be experimentally detected with digital quantum simulations.

\textit{Summary.}---%
We have investigated the mesons in QMBSs of a $\mathbb{Z}_2$ LGT.
By introducing the string representation, we express the wave function of each QMBS  as  an equal superposition of all string
bases with an identical string number and total string length.
We demonstrate that scar states with a small total string length are described by stable mesons, like conventional towers of QMBSs in LGTs,
while we find the nonmesonic excitations in the scar states with large total string length.
Furthermore, this  mesonic-nonmesonic crossover in QMBSs can be observed from the quench dynamics with two experimentally accessible initial states.
Our results bring new insights into QMBSs in LGTs and reveal that the nonmesonic states can also host ergodicity breaking in LGTs,
which can be experimentally verified with quantum simulators.

Here, the mesonic-nonmesonic crossover is reminiscent of the asymptotic freedom of quarks~\cite{PhysRevLett.30.1343,PhysRevLett.30.1346}.
The eigenenergy of the scar state $\ket{\Psi_{n,l}}$  becomes large when increasing the total string length $l$.
Thus, the nonmesonic excitation in large-$l$ QMBSs is in analogy with the asymptotic freedom of quarks
in the high-energy regime of quantum chromodynamics.
The conventional QMBSs are generated by local operators, which correspond to nonfractionalized excitations~\cite{doi:10.1146/annurev-conmatphys-031620-101617},
e.g., magnons in \textit{PXP}~\cite{PhysRevB.100.184312} and spin-$1$ \textit{XY} models~\cite{PhysRevLett.123.147201}, and $\eta$-pairs in Hubbard-like models~\cite{PhysRevB.102.075132,PhysRevB.102.085140}.
However, the nonmesonic excitations have some analogies with spinons~\cite{Fradkin2013}, which are fractionalized excitations and different from conventional QMBSs.
Therefore, our results also reveal that nonlocal generating operators may lead to fractionalized excitations in QMBSs,
providing an inspiration for studying the nontrivial excitation in QMBSs.
Here, nonlocal generating operators are necessary for the nonmesonic QMBSs,
since they can separate two matter particles of a meson.
However, it is still unclear which types of  generating operators can induce these nontrivial physics.
Another interesting issue is whether the above physics can be generalized to other gauge groups or high-dimensional LGTs~\cite{PhysRevB.38.2926,banuls2020simulating}.
These open questions deserve further study.

\begin{acknowledgments}
	\textit{Acknowledgments.}---%
	We thank Marcello Dalmonte, Zongping Gong, and Rui-Zhen Huang for insightful discussions and useful suggestions.
	This work is supported in part by:
	Nippon Telegraph and Telephone Corporation (NTT) Research,
	the Japan Science and Technology Agency (JST) [via
	the Quantum Leap Flagship Program (Q-LEAP), and
	the Moonshot R\&D Grant Number JPMJMS2061],
	the Asian Office of Aerospace Research and Development (AOARD) (via Grant No. FA2386-20-1-4069), and
	the Office of Naval Research (ONR) Global (via Grant No. N62909-23-1-2074).
\end{acknowledgments}

 \clearpage
\widetext

\begin{center}
	\subsection{\large Supplemental Material: \\
		\textit{Nonmesonic Quantum Many-Body Scars in a 1D Lattice Gauge Theory} }
\end{center}

\setcounter{equation}{0} \setcounter{figure}{0}
\setcounter{table}{0} \setcounter{page}{1} \setcounter{secnumdepth}{3} \makeatletter
\renewcommand{\theequation}{S\arabic{equation}}
\renewcommand{\thefigure}{S\arabic{figure}}
\renewcommand{\bibnumfmt}[1]{[S#1]}

\makeatletter
\def\@hangfrom@section#1#2#3{\@hangfrom{#1#2#3}}
\makeatother

\section{Quantum Many-Body Scars}

\subsection{Proof of $\hat H_K \ket {\Psi_{n,l}} = 0$}
In the main text, we show that the wave function 
\begin{align} 
	\ket {\Psi_{n,l}} = \mathcal{N}_{n,l}\sum P_{\{\mathcal{S}_{k_j,\ell_j }\}_n ^l} \ket{\{\mathcal{S}_{k_j,\ell_j }\}_n^l },
\end{align}
is an exact eigenstate of $\hat H$.
Here we present details for proving this result.
It is not difficult to find that $\hat H_E \ket {\Psi_{n,l}} =h(2l-L) \ket {\Psi_{n,l}}$,
so we only need to prove $\hat H_K \ket {\Psi_{n,l}} = 0$.
Since the action of $\hat H_K$ is increasing or reducing the total string length by one, while keeping $n$ invariant,  we have
\begin{align} 
	\hat H_K \ket {\Psi_{n,l}} = \sum c_{\{\mathcal{S}_{k'_j,\ell'_j }\}_n ^{l-1} } \ket{\{\mathcal{S}_{k'_j,\ell'_j }\}_n^{l-1}  } +\sum c_{\{\mathcal{S}_{k'_j,\ell'_j }\}_n^{l+1}  } \ket{\{\mathcal{S}_{k'_j,\ell'_j }\}_n^{l+1}  }.
\end{align}
Here, the factors have forms 
\begin{align} \nonumber
	c_{\{\mathcal{S}_{k'_j,\ell'_j }\}_n^{l-1} }  =  \mathcal{N}_{n,l}\sum  [&(1-\delta_{ k'_{1}+\ell'_{1}+1,k'_2 })(P_{\mathcal{S}_{k'_1,\ell'_1+1},\mathcal{S}_{k'_2,\ell'_2}...}+P_{\mathcal{S}_{k'_1,\ell'_1},\mathcal{S}_{k'_2-1,\ell'_2+1}...})\\ \nonumber
	+&(1-\delta_{k'_{2}+\ell'_{2}+1,k'_3 })(P_{...,\mathcal{S}_{k'_2,\ell'_2+1},\mathcal{S}_{k'_3,\ell'_3}...}+P_{...,\mathcal{S}_{k'_2,\ell'_2},\mathcal{S}_{k'_3-1,\ell'_3+1}...})+...\\ \nonumber	c_{\{\mathcal{S}_{k'_j,\ell'_j }\}_n^{l+1} }  =  \mathcal{N}_{n,l}\sum  [&(1-\delta_{ \ell'_{1},1 })(P_{\mathcal{S}_{k'_1-1,\ell'_1-1},\mathcal{S}_{k'_2,\ell'_2}...}+P_{\mathcal{S}_{k'_1,\ell'_1-1},\mathcal{S}_{k'_2,\ell'_2}...})\\
	+&(1-\delta_{\ell'_{2},1})(P_{...,\mathcal{S}_{k'_2-1,\ell'_2-1},\mathcal{S}_{k'_3,\ell'_3}...}+P_{...,\mathcal{S}_{k'_2,\ell'_2-1},\mathcal{S}_{k'_3,\ell'_3}...})+....
\end{align}
Since the parity satisfies $P_{\mathcal{S}_{k'_1,\ell'_1},...,\mathcal{S}_{k'_j,\ell'_1},...,\mathcal{S}_{k'_n,\ell'_n}}=\exp{(i\pi\sum_j k'_j)}$,
we have 
\begin{align} \nonumber
	&P_{...\mathcal{S}_{k'_{j-1},\ell'_{j-1}},\mathcal{S}_{k'_j,\ell'_j+1},\mathcal{S}_{k'_{j+1},\ell'_{j+1}}...}
	=-P_{...\mathcal{S}_{k'_{j-1},\ell'_{j-1}},\mathcal{S}_{k'_j,\ell'_j},\mathcal{S}_{k'_{j+1}-1,\ell'_{j+1}+1}...}\\
	&P_{...\mathcal{S}_{k'_{j-1},\ell'_{j-1}},\mathcal{S}_{k'_j-1,\ell'_j-1},\mathcal{S}_{k'_{j+1},\ell'_{j+1}}...}
	=-P_{...\mathcal{S}_{k'_{j-1},\ell'_{j-1}},\mathcal{S}_{k'_j,\ell'_j-1},\mathcal{S}_{k'_{j+1},\ell'_{j+1}}...}.
\end{align}
Therefore, $c_{\{\mathcal{S}_{k'_j,\ell'_j }\}_n^{l-1}} =c_{\{\mathcal{S}_{k'_j,\ell'_j }\}_n^{l+1}}=0$, i.e., $\hat H_K \ket {\Psi_{n,l}} = 0$.

\subsection{Proof of 	$\ket {\Psi_{n,n+m}}  = \mathcal{D}_{n,m}\hat L_m^\dagger \ket {\Psi_{n,n+m-1}}$}
Next we show the detail of proving Eq.~(8) in the main text, i.e,
\begin{align}
	\ket {\Psi_{n,n+m}}  = \mathcal{D}_{n,m}\hat L_m ^\dagger\ket {\Psi_{n,n+m-1}} ,
\end{align}
where $\mathcal{D}_{n,m}$ is a normalization factor, and 
\begin{align}
	\hat L^\dagger_m =\sum_{j}  \big(\sum_{k\leq m}\prod_{\ell \leq k}\hat {\mathcal{P}}^-_{j+\frac{1}{2}-\ell} \big)\hat \sigma^-_{j}\hat \tau^{\text{M}}_{j+\frac{1}{2}}\hat \sigma^+_{j+1}
\end{align}
It is not difficult to demonstrate
\begin{align}
	\big(\sum_{k\leq m}\prod_{\ell \leq k}\hat {\mathcal{P}}^-_{j+\frac{1}{2}-\ell} \big)\hat \sigma^-_{j}\hat \tau^{\text{M}}_{j+\frac{1}{2}}\hat \sigma^+_{j+1}\ket{\mathcal{S}_{k,\ell}} =
	\begin{cases}
		\ell\delta_{j,k+\ell} \ket{\mathcal{S}_{k,\ell+1}} \quad \ell \leq m  \\
		m \delta_{j,k+\ell}  \ket{\mathcal{S}_{k,\ell+1}} \quad \ell >m.
	\end{cases}	  
\end{align}
Thus, the action of $\hat L^\dagger_m $ is increasing the total string length of a basis without changing the parity and string number.
Therefore,
\begin{align}
	\hat L_m ^\dagger \ket {\Psi_{n,n+m-1}} = \sum \alpha_{\{\mathcal{S}_{k'_j,\ell'_j }\}_n ^{n+m} } \ket{\{\mathcal{S}_{k'_j,\ell'_j }\}_n^{n+m}  }.
\end{align}
For the wave function $\ket {\Psi_{n,n+m-1}} = \mathcal{N}_{n,n+m-1}\sum P_{\{\mathcal{S}_{k_j,\ell_j }\}_n^{n+m-1} } \ket{\{\mathcal{S}_{k_j,\ell_j }\}_n^{n+m-1}}$,
the length of each string satisfies $\ell_j \leq m$.
Hence, the factor has the form
\begin{align}
	\alpha_{\{\mathcal{S}_{k'_j,\ell'_j }\}_n ^{n+m} } = \mathcal{N}_{n,n+m-1}P_{\{\mathcal{S}_{k'_j,\ell'_j }\}_n ^{n+m}}[(1-\delta_{\ell'_1,1})(\ell'_1-1)+(1-\delta_{\ell'_2,1})(\ell'_2-1)+...+ (1-\delta_{\ell'_n,1})(\ell'_n-1)].
\end{align}
If $\ell'_j=1$, then $(1-\delta_{\ell'_j,1})(\ell'_j-1)=(\ell'_j-1)=0$, and if $\ell'_j \neq1$, then $(1-\delta_{\ell'_j,1})(\ell'_j-1)=(\ell'_j-1)$.
Thus 
\begin{align}
	\alpha_{\{\mathcal{S}_{k'_j,\ell'_j }\}_n ^{n+m} } = \mathcal{N}_{n,n+m-1}P_{\{\mathcal{S}_{k'_j,\ell'_j }\}_n ^{n+m}}\sum_{j=1}^{n}(\ell'_j-1)
	= \mathcal{N}_{n,n+m-1}P_{\{\mathcal{S}_{k'_j,\ell'_j }\}_n^{n+m}}(m-1) .
\end{align}
Therefore, we have
\begin{align}
	\hat L_m^\dagger \ket {\Psi_{n,n+m-1}} = (m-1)\mathcal{N}_{n,n+m-1}\sum P_{\{\mathcal{S}_{k'_j,\ell'_j }\}_n^{n+m}}\ket{\{\mathcal{S}_{k'_j,\ell'_j }\}_n^{n+m}  }
	=\frac{(m-1)\mathcal{N}_{n,n+m-1}}{\mathcal{N}_{n,n+m}}\ket {\Psi_{n,n+m}}.
\end{align}
That is, Eq.~(8) is proved, and the normalization factor satisfies
\begin{align}
	\mathcal{D}_{n,m}
	=\frac{\mathcal{N}_{n,n+m}}{(m-1)\mathcal{N}_{n,n+m-1}}.
\end{align}

\begin{figure}[t] \includegraphics[width=0.6\textwidth]{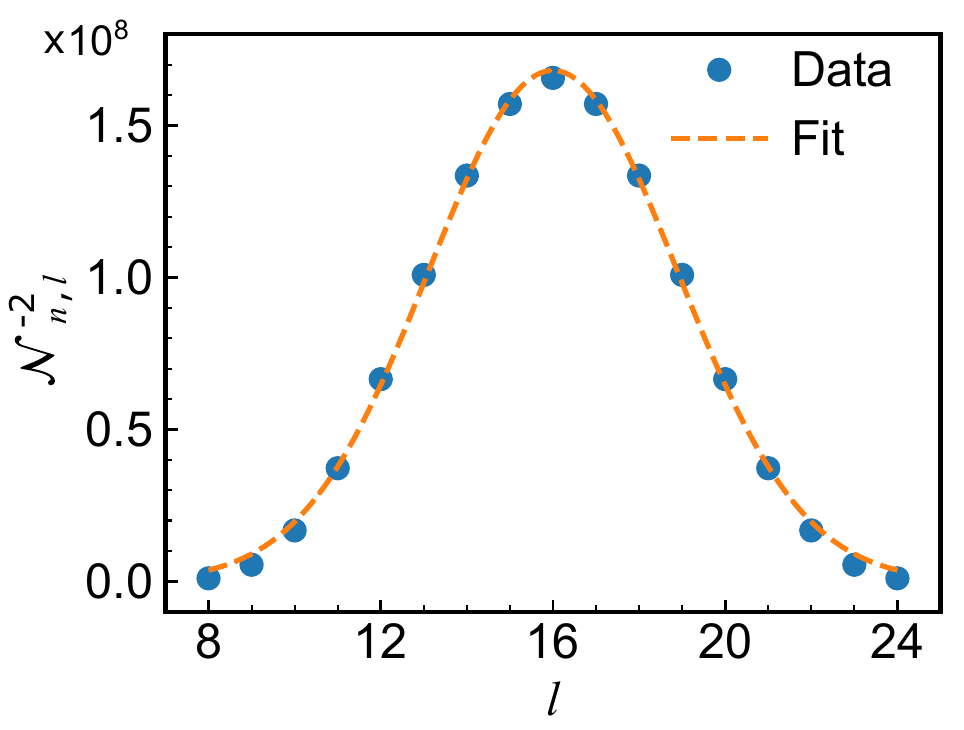}
	\caption{Distribution of $\mathcal{N}_{n,l}^{-2}$ for $L=32$ and $n=8$ (half filling). The orange dashed curve is a Gaussian fit.}
	\label{fig_s1}
\end{figure}

\section{Initial state}
Here we discuss the initial state $\ket{\psi_{2} }$ in Eq.~(10b) of the main text, where it reads
\begin{align} 
	\ket{\psi_{2} }&= \frac{1}{2^{L/2}} \sum_{n,l}\sum _{\{k_j,\ell_j\}} P_{\{\mathcal{S}_{k_j,\ell_j }\}_n^l } \ket{\{\mathcal{S}_{k_j,\ell_j }\}_{n}^{l} }
	=\sum_{n,l} \beta_{n,l}\ket{\Psi_{n,l}}.
\end{align}
The amplitude $\beta_{n,l}$ satisfies $\beta_{n,l} = 1/\mathcal{N}_{n,l} 2^{L/2}$, 
where $\mathcal{N}_{n,l} $ is the normalization factor defined in Eq.~(6) of the main text.
In addition, $\mathcal{N}_{n,l}^{-2} $ is the number of string bases for the scar state $\ket{\Psi_{n,l}}$,
and it can be obtained as 
\begin{align} 
	\mathcal{N}_{n,l}^{-2} =\binom{l-1}{n-1} \bigg[\binom{L-l-1}{n}+2\binom{L-l-1}{n-1}\bigg]+\binom{L-l-1}{n-1}\binom{l-1}{n},
\end{align}
where $\binom{\cdot}{\cdot}$ is the combinatorial number.
In Fig.~\ref{fig_s1}, we show the result of $\mathcal{N}_{n,l}^{-2}$ versus $l$ for $L=32$ and $n=8$ (half filling).
We can find that $\mathcal{N}_{n,l}^{-2}$  nearly satisfies a Gaussian distribution with the symmetric point at $l=L/2$.
Therefore, for the initial state $\ket{\psi_{2} }$ the nonmesonic scar states dominate.


\begin{thebibliography}{90}
\bibitem [{Str()}]{StringLength}%
	\BibitemOpen
	\bibinfo {title} {Due to the chiral symmetry in the Hamiltonian when
		neglecting $\hat H_\mu $, i.e., $\hat P \hat H \hat P^\dagger = -\hat H$ with
		$\hat P := \prod_{j}\hat \tau_{j+\frac{1}{2}}^y$, a scar state with total
		string length $l\geq L/2$ is equivalent to a scar state with total string
		length $(L-l)$. In addition, the eigenenergy of a scar state with $l=L/2$ is
		zero (the central part of the spectrum), corresponding to an
		infinite-temperature system. Therefore, we only need to consider the total
		string length $l\leq L/2$, and the large total string length means that $l$
		is $\approx L/2$.},\ \href@noop {} {\ }\BibitemShut {NoStop}%
\end{thebibliography}
\end{document}